\documentstyle[12pt]{article}
\begin{document}
\vspace*{0.7cm}
\begin{center}
{\Large   Inflationary Cosmology with a
 $R + \lambda~R_{\mu \nu}~R^{\mu \nu}/R$\\
Lagrangian}\\
\bigskip
{\bf E. Br\"{u}ning \footnote {Present address: Dept. of
Mathematics, University of Durban-Westville, 4000 Durban, SA.}
, D.H. Coule and \\
C. Xu \footnote {Present address: Dept. of Physics,
 University of T\"{u}bingen,
 Germany. Also on leave from Fudan
University, Shanghai, China. }
}
\bigskip
\\
  Dept. of Applied Mathematics.
University of Cape Town\\
 Rondebosch 7700, SA
\begin{abstract}
We consider an alternate fourth-order gravity Lagrangian
which is non-analytic in the
Ricci scalar, and apply it to a Robertson-Walker metric. We find
vacuum solutions which undergo power-law Inflation. Once matter
is introduced the theory behaves very much like ordinary General
Relativity; except
that the radiation evolution $a\sim t^{1/2}$ is not allowed since
it corresponds to $R=0$.
 We comment on the possibility of wormhole solutions
in such a theory.
\end{abstract}
\end{center}
\newpage
\section{Introduction}

Inflationary universe models can be constructed by adding higher order
corrections to the usual Einstein-Hilbert Lagrangian for gravity. The
first such model was constructed using the trace anomaly of
conformally coupled fields which is due to a term $\sim R^2logR/\mu$
($\mu$ = renormalization scale) in the Lagrangian[1]. However, typical
numbers and types of field in the early universe would not produce
enough inflation, and any inflationary phase would finish too
abruptly causing an inhomogeneous universe. \\

The model was modified by adding a term $\epsilon R^2$ to the Lagrangian
which results in a quasi-de Sitter expansion[2-4].
 The Hubble parameter decreases
slowly for large $\epsilon$ before going into an oscillation phase which
can reheat the universe[4,5].
The general Lagrangian so far
considered has the form \footnote { Some works use the equivalent
in 4 dimensions
$R+aR^2+bC^2$ Lagrangian, where $C$ is the Weyl tensor.}
\begin{equation}
{\cal L} = R+\epsilon R^2 +\beta R_{\mu \nu}R^{\mu \nu}
\end{equation}
The coefficients $\epsilon $ and $\beta$ have dimensions of [length]$^2$
and  (in order for stability or to prevent tachyonic solutions) certain signs
have to be taken. The $\beta$ term is only relevant for anisotropic models
since in the isotropic case it just alters the $\epsilon$ coefficient.
 New gravitational waves have also been found in such Lagrangians[6].

Various other Lagrangians such as $R+\epsilon R^2 +\beta R^3$ and
$exp(\lambda R)$ have been considered and generally give power law inflation
[7-9]: instead of exponential expansion there is typically an expansion law
$a\sim t^m$. So long as $m>1$ and expansion occurs for sufficient time  the
usual problems, which are solved by inflation,  can still be solved.

Some papers have appeared concerning the stability and solutions of
general Lagrangians $f(R)$[11-14].
It is sometimes claimed that such theories or
their quasi-de Sitter solutions are unstable [11,12] but at least
classically this is not the case. This was analyzed in detail in ref.[13]
for the $R+\epsilon R^2$ theory and also found to match ordinary gravity
as $\epsilon\rightarrow 0$. The so-called small perturbations considered
in ref.[11] are actually abnormal: they correspond to a large $dH/dt$
($H$=Hubble parameter). In analogy with a scalar field inflationary
model this $\dot{H}$ term  is like a velocity causing you to shoot up
the scalar potential - but it is eventually  damped [15].

A recent problem has arisen in the case of Bianchi type-I models for
the Lagrangian (1) for $\beta \neq  0$ so including the sign that
is stable in the absence of anisotropy. The model is found to
collapse due to the
anisotropy [16]: this was also found by one of us previously using the
equations presented in ref.[17] but not taken too seriously at the
time since it conflicted with ref.[18]. Note that this is more restrictive
than for the case of scalar field inflation, where spatial gradients can
cause collapse[19] but not simply anisotropy[20].

Instead in this paper we consider an alternative higher
order derivative theory, first proposed in ref.[21]
\begin{equation}
{\cal L }= R + \lambda  R_{\mu \nu}R^{\mu \nu}/R + \tau~R_{\alpha
\beta \gamma \delta}R^{\alpha \beta \gamma \delta}/R.
\end{equation}

Because the parameters are now dimensionless such a Lagrangian could
help incorporate higher derivative theories into variable $G$
ones (eg. Brans-Dicke[22]) or those involving phase transitions c.f.
induced gravity [23]. These terms in the Lagrangian
might result when quantum gravity
corrections are made. Although we are not aware
of them being invoked from string theory, which is usually taken to
give corrections of the form Lagrangian (1), this might change when
the theory is better developed or is superceded.

In any case such a non-analytic Lagrangian is interesting in its
own right. It is considerably more complex than Lagrangian (1)
and requires us
to look for solutions with $R\neq 0$. If it
is found to have vacuum
Inflationary solutions, it would be a further extension
to the generality of higher derivative theories in
having such  Inflationary behaviour. Since we would expect that such
corrections to the Lagrangian would be present in the early universe
,when quantum gravity effects were dominating, they could help
explain any inflationary epoch.

\section{ Field Equations} \vspace*{1mm}
We consider an action of the form
\begin{equation}
S = \int \left[ (R + \lambda  R_{\mu \nu}~R^{\mu \nu}/R + \tau
R_{\alpha \beta \gamma \delta}~R^{\alpha \beta \gamma
\delta}/R)~\sqrt{-g} + \frac{8 \pi
G}{c^{2}}~L_{m} \right]~d^{4}x~.
\end{equation}

The field equations follow from variations
of this action leading to
\begin{equation}
G^{\mu \nu} = G^{\mu \nu}_{E} + \lambda G^{\mu \nu}_{1} + \tau
G^{\mu \nu}_{2} = - \frac{8 \pi G}{c^{2}}~T^{\mu \nu} ,
\end{equation}
where
\begin{equation} G^{\mu \nu}_{E} = R^{\mu \nu} - \mbox{$\frac{1}{2}$}~g^{\mu
\nu}~R ,
\end{equation}
\begin{eqnarray}
G^{\mu \nu}_{1} & = & g^{\mu \nu}~(R^{\sigma \tau}/R)_{;
\sigma \tau} + (R^{\mu \nu}/R)_{; \tau}{}^{\tau} - g^{\mu
\nu}~(R_{\alpha \beta}~R^{\alpha \beta}/R^{2})_{;
\tau}{}^{\tau} \nonumber \\
& - & [(R^{\mu \sigma}/R)_{;}{}^{\nu}{}_{\sigma} + (R^{\nu
\sigma}/R)_{;}{}^{\mu}{}_{\sigma}]
 +  (R_{\alpha \beta}~R^{\alpha \beta}/R^{2})_{;}{}^{\mu
\nu} \nonumber \\
& + & R^{\mu \nu}~(R_{\alpha \beta}~R^{\alpha \beta}/R^{2}) - 2
R^{\mu \sigma}~R^{\nu}{}_{\sigma}/R +
\mbox{$\frac{1}{2}$}~g^{\mu \nu}~R_{\alpha \beta}~R^{\alpha
\beta}/R, \\[2ex]
G^{\mu \nu}_{2} & = & 4 (R^{\mu \sigma \nu \tau}/R)_{;
\sigma \tau} - g^{\mu \nu}~(R_{\alpha \beta \gamma
\delta}~R^{\alpha \beta \gamma \delta}/R^{2})_{;
\tau}{}^{\tau} \nonumber \\
& + & (R_{\alpha \beta \gamma \delta}~R^{\alpha
\beta \gamma \delta}/R^{2})^{\mu \nu} \nonumber -
  \frac{2}{R}~R^{\mu}{}_{\sigma \tau
\rho}~R^{\nu \sigma \tau \rho}/R \nonumber \\
& + & R^{\mu \nu}~R_{\alpha \beta
\gamma \delta}~R^{\alpha \beta \gamma \delta}/R^{2} +
\mbox{$\frac{1}{2}$}~g^{\mu \nu}~R_{\alpha \beta \gamma
\delta}~R^{\alpha \beta \gamma \delta}/R~.
\end{eqnarray}
Since $R^{\mu \nu}$ and $R^{\alpha \beta \gamma \delta}$
include second order derivatives of the metric, the field
equations are fourth-order partial differential equations,
which is the same order as the normal higher order
gravitational theory.

Before considering a Robertson-Walker metric  we can find the
equation for a maximally symmetric spacetime. Such a spacetime has
\begin{equation}
R_{\mu \nu} = \mbox {$\frac{1}{4}$}~g_{\mu \nu}~R,
\end{equation}
\begin{equation}
R_{\alpha \beta \gamma \delta} =
- R~(g_{\alpha \gamma}~g_{\beta \delta} -
g_{\alpha \delta}~g_{\gamma \beta})~/12.
\end{equation}
Substituting (8) and (9) into (6) and (7), respectively, and
since the covariant derivatives of $g_{\mu \nu}$ are equal to
zero, we have
\begin{eqnarray}
G^{\mu \nu}_{1} & = & R^{\mu \nu}~R_{\alpha \beta}~R^{\alpha
\beta}/R^{2} - 2~R^{\mu \sigma}~R^{\nu}_{\sigma}/R \nonumber \\
& + & \mbox{$\frac{1}{2}$}~g^{\mu \nu}~R_{\alpha \beta}~R^{\alpha
\beta}/R  =  \mbox{$\frac{1}{16}$}~g^{\mu \nu}~R, \\[2ex]
G^{\mu \nu}_{2} & = & - 2~R^{\mu}{}_{\sigma
\tau \rho}~R^{\nu \sigma \tau \rho}/R + R^{\mu
\nu}~R_{\alpha \beta \gamma \delta}~R^{\alpha \beta
\gamma \delta}/R^{2} \nonumber \\
& + & \mbox{$\frac{1}{2}$}~g^{\mu \nu}~R_{\alpha
\beta \gamma \delta}~R^{\alpha \beta \gamma \delta}/R
 =  \mbox{$\frac{1}{24}$}~R~g^{\mu \nu}.
\end{eqnarray}
Substituting (10) and (11) into (4), we find
\begin{equation}
\mbox{$\frac{1}{4}$}~g^{\mu \nu}~R(1 -
\lambda/4 - \tau/6) =  8 \pi G/c^{2}~T^{\mu \nu}~,
\end{equation}
\begin{equation}
R = \frac {8 \pi G}{c^2} T / (1 - \lambda/4 - \tau/6)
\end{equation}
which is just the Einstein's field equation in a maximally
symmetric space except that the `` Gravitational constant''
 is $G(1 - \lambda/4 -
\tau/6)^{-1}$. When this constant is $\infty$ we can perhaps expect
a de-Sitter solution even in the vacuum case i.e. when $T=0$.\footnote{
As pointed out by the referee this argument is somewhat unsatisfactory,
but does give a heuristic argument as to why de-Sitter solution are
present.}
\section{Robertson-Walker Case }

 For simplicity we consider only
the Ricci squared term  so setting $\tau=0$.
 We also take a flat $k=0$ Robertson-Walker metric:
\begin{equation}
ds^2=-dt^2+a^2\left (dr^2+r^2(d\theta^2+sin^2\theta d\phi^2)\right )
\end{equation}
We use the conventions of Weinberg[24] such that the Ricci-scalar $R$ is
related to the Hubble parameter $H$ as,
\begin{equation}
R=-12H^2-6\dot{H}
\end{equation}
With this metric the field equations give just two independent
equations. The trace
equation
\begin{eqnarray}
\left (\frac{2}{3}-\frac{1}{\lambda}\right )R +24H^2 +396 \frac{H^4}{R}
+2016\frac{H^6}{R^2} +96H^3\frac{\dot{R}}{R^2}+936H^5\frac{\dot{R}}{R^3}
\nonumber\\
+216H^4\frac{\dot{R}^2}{R^4}-72H^4\frac{\ddot{R}}{R^3}
=\frac{8\pi G}{\lambda}\left ( \rho-3p\right ) \equiv \kappa _1
\end{eqnarray}
and the $(0,0)$ component equation
\begin{eqnarray}
\left (6+\frac{3}{\lambda}\right ) H^2+60\frac{H^4}{R}
-252 \frac{H^6}{R^2}+24H^3\frac{\dot{R}}{R^2}-72H^5\frac{\dot{R}}{R^3}
\nonumber \\=\frac{8\pi G}{\lambda}\rho \equiv \kappa_2
\end{eqnarray}
 We have also included a perfect fluid matter source with equation of
state $p=(\alpha-1)\rho$  ($p$ and $\rho$ are the pressure and
energy density respectively).
The conservation equation is given by
\begin{equation}
\dot{\rho}=-3\alpha H\rho \Rightarrow \rho=\frac{\rho_0}{a^{3\alpha}}
\end{equation}
A few well known equations of state are $\alpha=4/3$ (radiation);
$\alpha=1$ (matter) and $\alpha=0$ (cosmological constant)

Scaling properties of equations (16) and (17) with respect to changes
in the time scale suggest we look for solutions
for which $R$ and $H^2$ are proportional to each other
i.e. to look for solutions for which
\begin{equation}
\sigma =-\frac{H^2}{R}
\end{equation}
where $\sigma$ is a constant to be determined.
The $-ve$ sign is chosen to keep $\sigma +ve$ in our notation, a plus sign
would require a large  velocity $\dot{H}$  and would not be expected
to be inflationary cf. eq.(15).
Using the relation eq.(15) and  the ansatz eq.(19)
gives for the Hubble parameter
\begin{equation}
H=\frac{H_0}{1+H_0(2-1/6\sigma)t}
\end{equation}

Since  $a=a_0exp (\int Hdt)$ this leads
to solutions of the form
\begin{equation}
a(t)=a_0\left (1+H_0(2-\frac{1}{6\sigma} )t\right )^
{\frac{6\sigma}{12\sigma -1}}
\end{equation}
with $H_o$ and $a_o$ the initial values of the Hubble parameter and
scale factor respectively.
Eq.(21) therefore includes solutions like $a\sim t^m$ with
$m=6\sigma/(12\sigma-1)$.

If $(2-1/6\sigma) =0$ then simply $H=H_0$ and
\begin{equation}
a=a_0exp(tH_0)
\end{equation}

 This  usual de-Sitter solution corresponds as expected to
$\sigma=1/12$ when $ R=-12H^2$.

The ansatz when substituted into the  Trace equation and the
$(0,0)$ component equation , gives respectively

\begin{equation}
R\left (12\sigma^2+4\sigma -\frac{2}{3} +\frac{1}{\lambda}\right )=-\kappa_1
\end{equation}
\begin{equation}
3\sigma R \left (12\sigma^2+4\sigma-\frac{2}{3}+\frac{1}{\lambda}\right )=
-\kappa_2
\end{equation}
{\bf Vacuum Case}

We first consider the absence of matter i.e. $\kappa_1=\kappa_2=0$.
Since we require $R\neq0$ we set the $\sigma$ expression to be zero.
\begin{equation}
12\sigma^2+4\sigma-\frac{2}{3}+\frac{1}{\lambda}=0
\end{equation}
with solution
\begin{equation}
\sigma=-\frac{1}{6}+\frac{\sqrt 3}{6}\left ( 1-\frac{1}{\lambda}
\right ) ^{\frac{1}{2}}
\end{equation}

The other solution of the quadratic equation can be ignored since we
require $\sigma +ve$. This gives for $m$
\begin{equation}
m=\frac{-1+\sqrt{ 3} \left (1-\frac{1}{\lambda} \right )^{1/2}}
{-3+2\sqrt{ 3} \left (1-\frac{1}{\lambda} \right )^{1/2} }
\end{equation}

\underline{$\lambda $ positive:}\\
Real i.e. non oscillatory solutions,
 require $\lambda >1$ ; the de Sitter solution $m
\rightarrow \infty $ occurs when  $3=2\surd 3 (1-1/\lambda)^{1/2}
\Rightarrow  \lambda=4$. We can plot the behaviour of $m$ against
$\lambda$ in Fig.(1) . For $\lambda >4$ we have power law
inflationary behaviour
with $m\simeq 1.6$ as $\lambda \rightarrow \infty$.

We consider the stability of these solutions
with respect to linearized perturbations in the appendix. We find that
they are stable to such perturbations.\\
 For $1<\lambda<4$ the
universe is contracting i.e $m<0$. There is also an oscillation region
for $0<\lambda\leq1$ where $m$ becomes complex: but since we are primarily
interested in inflationary behaviour we do not consider it further.

\underline{$\lambda $ negative:}\\
 As $\lambda $ decreases  the expansion rate $m$ increases with $m\simeq
1.6$ as $\lambda \rightarrow -\infty$ . To get Inflationary behaviour
(i.e. $m>1$) requires $\lambda<-3$. Again these inflationary solutions
are stable to linearized perturbations.

{\bf Matter Case}

We now consider the addition of an energy density such that
\begin{equation}
\rho =\frac{\rho_0}{a^n}
\end{equation}
We are setting $n=3\alpha$ cf. eq.(18). Sticking with our same ansatz the
trace equation and $(0,0)$ equations gives respectively,
\begin{equation}
\left (\frac{2}{3}-\frac{1}{\lambda}-4\sigma-12\sigma ^2\right ) R
=\frac{8\pi G}{c^2\lambda}\frac{\rho_0(4-3\alpha)}{a^n}
\end{equation}
\begin{equation}
\left ( \frac{2}{3}-\frac{1}{\lambda}-4\sigma-12\sigma ^2 \right )R
=\frac{8\pi G}{c^2\lambda}\frac{\rho_0}{3\sigma a^n}
\end{equation}
 Equating the two gives
\begin{equation}
\frac{1}{3\sigma}=4-3\alpha
\end{equation}
The solution has the expansion behaviour $m=2/n = 2/3\alpha$
which is the usual relationship in standard General Relativity.

Since for the solution eq.(21) we have
\begin{equation}
H(t)=H_0\left (\frac{a_0}{a(t)}\right )^{1/m}
\end{equation}

the Ricci scalar can be written as
\begin{equation}
R=-\frac{1}{\sigma}H_0^2a_0^{2/m}a(t)^{-2/m}
\end{equation}
where the initial values $H_o, a_o $ and $\rho_o$ are related by eq.(30).\\
We have to  avoid $R=0$ which occurs when $1/\sigma =0$ i.e.
$4\neq 3\alpha $ or $n\neq 4$. This is the equation of state for
radiation whose scale factor behaves as $a\sim t^{1/2}$. This is
not allowed in this theory but a correction $a\sim t^{1/2+
\delta }$  with $\delta$ small is.

{\bf Cosmological constant}\\
This is a special case of the previously example with $n=0$ i.e. $\rho
=\rho_0$. Since we expect de-Sitter solutions we
 consider  $H={\rm constant}$. From the Trace and $(0,0)$ equations, we get
\begin{equation}
12\left ( \frac{1}{\lambda}-\frac{1}{4}\right )H^2=\kappa_1
\end{equation}
\begin{equation}
3\left ( \frac{1}{\lambda}-\frac{1}{4}\right )H^2 =\kappa_2
\end{equation}
Therefore
\begin{equation}
4=\frac{\kappa_1}{\kappa_2}=\left ( 1-\frac{3p}{\rho} \right )
\Rightarrow p=-\rho
\end{equation}

A de-Sitter solution occurs for the equation of  state $p=-\rho$
or equivalently a cosmological constant $\Lambda \sim 8\pi G
\rho$

Because  $H$ should be real the value of $\lambda$ is
constrained, from eq.(35)
\begin{equation}
\left ( 1-\frac{\lambda}{4}\right )H^2=\frac{8\pi G}{3c^2}\rho
\end{equation}
therefore $\lambda<4$ is required for such solutions.

\section{ Discussion and Conclusions}
We have found that provided the value of $|\lambda|$ is sufficiently
large the alternative higher order Lagrangian can have power law
inflationary solutions, which are stable
to linearized perturbations. Once matter is introduced, perhaps by a
quantum process (c.f. particle production
see e.g ref.[25]) in the early universe,
 the usual behaviour of General Relativity(GR)results. This is
provided the matter is allowed to dominate while the tendency
of the vacuum inflationary solution is to dilute any matter present.
If instead quantum
process proceed
abruptly it  might cause the inflationary phase to  rapidly stop-a
 graceful exit problem. Until the coupling to matter was known it
would be difficult to calculate the effects of particle
creation and how such a scenario could proceed: whether such
matter creation
can cause the  inflationary behaviour to
gradually switch over to a normal expansion $a\sim t^p$ with $p<1$.
In any case, there have been claims that
the usual inflationary models are unstable to quantum processes[26].
 If this is the case then not
having a ``classical" ending to inflation need not be a drawback: inflation
would end due to quantum instabilities and a mechanism
to enable a switch to a classical
non-inflationary phase is provided. A weak quantum instability would
therefore appear to
be a necessity,in
this, and similar power law inflationary models-weak in the sense that
there is still a quantum
probability  of sufficient inflation occurring.
\\
These quantum processes would also be required to heat the universe since
there is no oscillating field present:
this is also a problem  with scalar field power law
inflationary models where the field potential
is too shallow. If such processes are not present
or cause the inflationary expansion to prematurely stop then a second
period of inflation would be required. The first period of inflation
\footnote{Or at least enabled the universe to wait,
without
recollapsing prematurely c.f. ref.[27], till a lower
energy scale inflationary mechanism could dominate.}
would still have contributed to the flatness and horizon problem so
that the second inflationary stage (presumably due to a matter field)
could be shorter than the usual $\sim 70$ e-foldings.\\
  There
is the problem that the radiation expansion $a\sim t^{1/2}$ cannot quite
occur (this would set the Ricci scalar zero). This might conflict with
the usual predictions of nucleosynthesis and so require that $\lambda$
be smaller.\footnote{We have not yet calculated how $\lambda$ affects the
nucleosynthesis rates and if any possible differences disappear as $\lambda
\rightarrow 0$.}

We could still envisage that during the early epoch of the universe the
 value of $\lambda$ is large: so giving inflation. As
quantum gravity effects weaken due to the expansion of the universe the
value of $\lambda$ reduces, so effectively agreeing with GR later.
We would hope to show that as $\lambda\rightarrow 0$ we
would recover or agree with the results of GR. In this
case $\lambda$ varies like a ``running
coupling constant'' in particle physics due
to the relevant energy scale present.

 An alternative scenario would be
for $\lambda$ to be simulated by the effect of extremely  highly
energetic matter e.g. a string epoch. $\lambda$ could then change
suddenly rather like the induced gravity model c.f ref.[23]
 due to the different
behaviour of the matter fields present (in the
string epoch this could correspond to a phase transition ).
 This would require us to
understand the matter fields present, and their coupling to the
higher derivative terms, before we would know if they could set
$\lambda$ small. Note that the field equations have been derived
under the assumption that $\lambda$ is independent of time. If this
assumption was dropped then the field equations would be more
 complicated.

 Because the
effective Newton's constant $G_{eff}$
can change sign there is also the chance of
finding wormhole solutions. This is similar to the wormhole found
in the case of a conformally coupled scalar field which has $G_{eff}
<0$ in its vicinity[28,29]. Wormholes can be found in the usual
higher derivative  Lagrangian (1) if we take the ``wrong sign''
of epsilon[30,31]: this
is related to the bounce (avoidance of the singularity)
in such models for spatial curvature $k=-1$[31].\\
 In order to find wormholes (or bounces) in the Lagrangian
presented here we would have to keep the spatial curvature $k$ in the
equations. This would be rather more complicated as would the
 case of Bianchi
models: these could perhaps
be more realistically tackled using algebraic computing packages.

In conclusion we have obtained vacuum inflationary solutions for
a non analytic higher order
Lagrangian.
 This Lagrangian  might contain certain  aspects of
the ``correct quantum gravity'' theory at high energy densities.
 Such solutions are  further evidence that De Sitter
or power law inflationary
solutions are a general feature of higher order
gravitational theories.
\\

{\bf Acknowledgement}
C.Xu would like to thank Prof. P. Havas for his enlightening
discussions on various aspects of the alternate gravity theory.
We are grateful to the referees for many useful comments and
suggestions which are reflected in this
revised version.\\

{\bf Appendix}\\
The question of stability of the solutions is clearly of importance.
However, a comprehensive discussion of this topic would include
such concepts as e.g. linearized, weak, asymptotic, and Liapunov stability.
This would entail considerable work beyond the
scope of this paper, so we limit ourselves to just an
analysis of the linearized stability of the vacuum equations.
For a general reference see eg.[32].\\
 First
consider the $(0,0)$ eq.(17) rewritten in the form
\begin{equation}
\dot{R}=\frac{1}{(1-3H^2/R)}\left ( \frac{21}{2}H^3-\frac{5}{2}HR
-\frac{1}{8}(2+\frac{1}{\lambda})\frac{R^2}{H}\right ) \equiv F_2(H,R)
\end{equation}
together with the expression for the Ricci scalar
\begin{equation}
\dot{H}=-2H^2-R/6 \equiv F_1(H,R)
\end{equation}
We consider perturbations $h$ and $r$ around the known solutions
such that now
\begin{equation}
H=\overline{H}+h \;\;\;\;  R=\overline{R}+r
\end{equation}
where $\overline{H}$ and $\overline{R}$ are the previously
found solutions.
The linearized equations for the perturbations are given by
\begin{equation}
\dot{h} =\frac{\partial F_1(H,R)}{\partial H}h +\frac{\partial F_1(H,R)}
{\partial R}r
\end{equation}
\begin{equation}
\dot{r}=\frac{\partial F_2(H,R)}{\partial H}h+\frac{\partial F_2(H,R)}
{\partial R}r
\end{equation}
Calculating the partial derivatives eventually leads to
\begin{equation}
\dot{h}=-4h\overline{H}-r/6
\end{equation}
\begin{equation}
\dot{r}=\beta_1h\overline{H}^2+\beta_2r\overline{H}
\end{equation}
where
\begin{equation}
\overline{H}^{-1}=H_0^{-1}+\gamma t\;,\;\;  \gamma\equiv (2-1/6\sigma)
\geq0
\end{equation}
and
\begin{equation}
\beta_1=\frac{1}{(1+3\sigma)^2}\left (18+18\sigma
+\frac{5}{\sigma}+\frac{1}{3\sigma^2}\right )
\end{equation}
\begin{equation}
\beta_2=\frac{1}{(1+3\sigma)^2}\left ( -\frac{1}{2}
-\frac{45}{2}\sigma-45\sigma^2  +\frac{2}{3\sigma}
\right ).
\end{equation}
In deriving $\beta_1,\beta_2$ we have made use of eq.(25) to express $\lambda$
in terms of $\sigma$.
Scaling to a new time parameter $\tau$ such that $d\tau/dt=\gamma$
and renaming $u(\tau)=h(t)$ and $v(\tau)=r(t)$ gives
\begin{equation}
\dot{u}(\tau)=\frac{1}{\gamma}
\left (-\frac{4u}{ \tau}-\frac{v}{6}\right )
\end{equation}
\begin{equation}
\dot{v}(\tau)=\frac{1}{\gamma}\left (
\frac{\beta_1 u}{\tau ^2}+\frac{\beta_2 v
}{\tau} \right )
\end{equation}
We have to find solutions with $\tau(\equiv
\overline{H}^{-1})\geq \tau_0 =H_0^{-1}$, with
initial conditions $u(\tau _0)\equiv u_0$ and $v(\tau_0)=v_0$.\\
The nature of the equations suggests we look for solutions of the
form
\begin{equation}
u=u_0 \tau^{-\alpha} \;\;\;\;\; v=v_0 \tau ^{-\alpha -1}
\end{equation}

When substituted into the equations this ansatz
leads to the linear
system for $u_0$,$v_0$
\begin{equation}
(4-\alpha \gamma)u_0 +v_0/6 =0
\end{equation}
\begin{equation}
\beta_1u_0+\left (\beta_2+(\alpha+1)\gamma \right )
v_0=0
\end{equation}
Non-trivial solutions require that the determinant vanish
i.e.
\begin{equation}
(4-\alpha\gamma)(\beta_2+(\alpha+1)\gamma)-\beta_1/6=0
\end{equation}
or
\begin{equation}
\alpha^2-\left (\frac{4-\beta_2}{\gamma}-1 \right )\alpha
-\left (\frac{4\beta_2-\beta_1/6}{\gamma^2}+\frac{4}{\gamma}
\right )=0
\end{equation}
with solutions
\begin{equation}
\alpha_{1,2}=\frac{1}{2}\left [\frac{4-\beta_2}{\gamma}-1 \right ]
\pm \sqrt{\frac{1}{4}\left [\frac{4-\beta_2}{\gamma}-1\right ]^2
+\frac{4\beta_2-\beta_1/6}{\gamma^2}+\frac{4}{\gamma}}\;\;,
\end{equation}
which can be written in the form (with obvious definitions for
$A,B$)
\begin{equation}
\alpha_{1,2}=\frac{A}{\gamma}\pm \frac{1}{\gamma}\sqrt {A^2+B}.
\end{equation}
For stability we require both $\alpha_1$ and $\alpha_2$ $>0$
this occurs if the following inequalities are satisfied (recall
$\gamma\geq 0$):
\begin{equation}
A>0 \;,\; B<0\;,\; A^2+B>0
\end{equation}
Substituting for $\beta_1$,$\beta_2$ allows these inequalities to
be written as
\begin{equation}
0<\frac{1}{(1+3\sigma)^2}\left ( \frac{7}{2}
+36\sigma+63\sigma^2 -\frac{1}{2\sigma}
\right )
\end{equation}
\begin{equation}
0<\frac{1}{(1+3\sigma)^2}\left (1+51\sigma+108\sigma^2+\frac{1}{18\sigma^2}
-\frac{7}{6\sigma}\right)
\end{equation}
\begin{equation}
0<(4-\beta_2-\gamma)^2+\frac{1}{(1+3\sigma)^2}\left (-4-204\sigma
-432\sigma^2+\frac{14}{3\sigma}-\frac{2}{9\sigma^2}\right )\;\;.
\end{equation}
The value of $\sigma$ when
inflationary behaviour occurs i.e. $\lambda>4$ is constrained in the
region
\begin{equation}
\frac{1}{12}\leq \sigma\leq \left (
\frac{1}{6}\sqrt{3}-\frac{1}{6}\right ) \simeq 0.122
\end{equation}
where we have again used eq.(26) to relate $\sigma$ to $\lambda$.
  For $\lambda <-3$ the corresponding expression is
\begin{equation}
\frac{1}{6}(\sqrt{3}-1)\leq\sigma\leq \frac{1}{6}
\end{equation}
With these values of $\sigma$ the inequalities are satisfied.
The first two are easily satisfied while the third one has been
checked numerically. In fact, the restriction to the inflationary
regime is not strictly necessary since the inequalities are only
violated in the oscillatory region $0<\lambda\leq1$.\\
 The
values of $\alpha_1$, $\alpha_2$ are therefore positive and the
perturbations decay  as $t$ or $\tau \rightarrow \infty$. Let us
represent the solutions for the
perturbations as $\omega(\alpha_1)$,$\;\omega(\alpha_2)$
, where $\omega\equiv \left ( \begin{array}{c} u\\v \end{array}
\right )$. These two independent solution of $\omega$ will span
the solution space of eq.(48,49), which is a two dimensional
vector space. This
is in contrast with the solutions (which
can be called $\Omega\equiv \left (
\begin{array}{c} \overline{H}\\
\overline{R} \end{array} \right )$) which go as $\overline{H}=1/\tau$,
$\overline{R}=-1/(\sigma \tau^2)$ (see eq.(19).\\
 In order
for the perturbations not to dominate they must decay
more rapidly  than their corresponding solutions i.e. we
require $\alpha>1$. Because  $\alpha_2$ is
less than $\alpha_1$, $\omega(\alpha_2)$ will give the dominant
behaviour  as $\tau\rightarrow \infty$.\\
Now $\alpha_2$ is given by
\begin{equation}
\alpha_2=\frac{A}{\gamma}-\frac{1}{\gamma}\sqrt{A^2+B}.
\end{equation}
Numerically it can be shown that $(A^2+B)\simeq 0.2$ so approximately
\begin{equation}
\alpha_2\simeq \frac{A}{\gamma}\equiv \frac{1}{2\gamma(1+3\sigma)^2}
\left (\frac{7}{2}+36\sigma+63\sigma^2-\frac{1}{2\sigma}\right)
-\frac{0.2}{\gamma}
\end{equation}
For $\sigma$ in the inflationary range ($\lambda >4$) we obtain
\begin{equation}
 \sim 1.4<\alpha_2\leq \infty
\end{equation}
(For $\lambda<-3$ the corresponding expression is $1.4<\alpha_2<1.8$.)\\
We can therefore conclude that since $\alpha_2>1$ the perturbations
$\omega(\alpha_2)$
will decay faster than their corresponding solutions $\Omega$.\\
Again this restriction to the inflationary range was not strictly
necessary except
for the oscillatory region.
Although we do not try to
analyze this region further, we would also hope that
for stability in this region where
$\alpha$ is complex, the real part i.e both $\alpha_{real}^{1,2}$
are  positive.\\
We would also like to show at some stage
 that the inclusion of matter is likewise
stable. Although we are fairly
confident that this will be the case, since more rapidly expanding
solutions are usually more
prone to instability, it would require extensive further work
to be certain.\\
The trace equation should also be considered but it can be shown
that it  only imposes restrictions on the initial values
$u_0$,$v_0$ and so does not add further to the stability analysis.
We can therefore conclude that the inflationary solutions are stable,
at least to linearized perturbations. We can be fairly confident that
other solutions (i.e oscillatory, matter dominated) will also be stable
without explicitly doing each case.

\newpage
{\bf REFERENCES}

\begin{enumerate}
\item A.A. Starobinsky, Phys. Lett. 91B (1980) p.99.\\
see also: Y. Hosotani, M. Nikolic and S. Rudaz, Phys. Rev. D 34 (1986)
p.627.
\item A.A. Starobinsky, Sov. Astron. Lett. 10 (1984) p.135.
\item L.A. Kofman, A.D. Linde and A.A. Starobinsky, Phys. Lett.
157 B (1985) p.361.
\item M.B. Miji\'{c}, M.S. Morris, and W-M Suen, {\it Phys. Rev.}
{\bf D34}, 2934 (1986).
\item P.A. Anderson and W.Suen, Phys. Rev. D 35 (1987) p.2940.
\item M.S. Madsen, Class. Quan. Grav. 7 (1990) p.87.\\
C. Xu and G.F.R. Ellis, Class. Quan. Grav. 8 (1991) p.1747.
\item J.D. Barrow and S. Cotsakis, Phys. Lett. B 214 (1988) p.515.
\item A.L. Berkin and K. Maeda, Phys. Lett. B 245 (1990) p.348.
\item A.R. Liddle and F. Mellor, Gen. Rel. and Grav. 24 (1992) p.897.
\item A.A. Coley and R.K. Tavakol, Gen. Rel. and Grav. 24 (1992) p.835.
\item W.M. Suen, Phys. Rev. Lett. 62 (1989) p.2217.\\
Phys. Rev. D 40 (1989) p.315.
\item J.Z. Simon, Phys. Rev. D 45 (1992) p.1953.
\item J.D. Barrow and A.C. Ottewill, J. Phys. A 16 (1983) p.2757.
M.S. Madsen and J.D. Barrow, Nucl. Phys. B 323 (1989) p.242.
\item M.S. Madsen and R.J. Low, Phys. Lett. B 241 (1990) p.207.
\item D.H. Coule and M.S. Madsen, Phys. Lett. B 226 (1989) p.31.
\item A.L. Berkin, Phys. Rev. D 44 (1991) p.1020.
\item K. Tomita, T. Azuma and H. Nariai, Prog. of Theor. Phys. 60
(1978) p.403.
\item H.J. Schmidt, Class. Quant. Grav. 5 (1988) p.233.
\item M.Y. Khlopov, B.A. Malomed and Y.B. Zeldovich, Mon. Not. R. Ast.
Soc. 215 (1985) p.575.
\item R. Wald, Phys. Rev. D 28 (1983) p.2118.
\item C. Xu and G.F.R. Ellis, Found. of Phys. Lett. 5 (1992) p.365.
\item C. Brans and C.H. Dicke, Phys. Rev. 24 (1961) p.925.
\item A. Zee, Phys. Rev. Lett. 42 (1979) p.417.
\item S. Weinberg, {\it Gravitation and Cosmology} (Wiley,
New York, 1972).
\item N.D. Birrell and P.C.W. Davies, \\
{\it Quantum fields in curved space}
(Cambridge University Press, 1982).
\item P. Mazur and E. Mottola, Nucl. Phys. B 278 (1986) p.694.\\
and references therein.
\item J.D. Barrow, Nucl. Phys. B 296 (1988) p.697.
\item J.J. Halliwell and R. Laflamme, Class. Quant. Grav. 6 (1989) p.1839.
\item Y. Verbin and A. Davidson, Nucl. Phys. B 339 (1990) p.545.
\item H. Fukutaka, K. Ghoroku and K. Tanaka, Phys. Lett. B 222 (1989) p.191.
\item D.H. Coule, Class. Quant. Grav. 10 (1993) p.L25.
\item M.W. Hirsch and S. Smale, {\it Differential equations,
Dynamical systems and Linear Algebra} (Academic Press, 1974).\\
R. Grimshaw, {\it
 Nonlinear Ordinary Differential Equations} (Blackwell Scientific,
Oxford, 1990).
\end{enumerate}
{\bf Figures}\\
Fig.1\\
Plot of the expansion coefficient $m$ against the parameter $\lambda$.
Power law inflation occurs for $\lambda\geq 4$ and $\lambda\leq -3$.
\end{document}